\documentclass[twocolumn,tighten,twocolappendix	]{aastex63}

\usepackage{times,natbib,graphicx,amsmath,multirow,xspace}
\usepackage{appendix}
\usepackage{xcolor}
\usepackage{lineno}

\newcommand{\straycats}{\texttt{StrayCats}\xspace}

\newcommand{\nustar}{\textit{NuSTAR}\xspace}
\newcommand{\maxi}{\textit{MAXI}\xspace}
\newcommand{\swift}{\textit{Swift}\xspace}

\shorttitle{StrayCats II}
\shortauthors{Ludlam et al.}

\begin{document}

\title{\textit{StrayCats II}: An Updated Catalog of NuSTAR Stray Light Observations}

\correspondingauthor{R.~M.~Ludlam}
\email{rmludlam@caltech.edu}
\author[0000-0002-8961-939X]{R.~M.~Ludlam}\thanks{NASA Einstein Fellow}
\affiliation{Cahill Center for Astronomy and Astrophysics, California Institute of Technology, Pasadena, CA 91125, USA}
\author[0000-0002-1984-2932]{B.~W.~Grefenstette}
\author[0000-0002-4024-6967]{M.~C.~Brumback}
\affiliation{Cahill Center for Astronomy and Astrophysics, California Institute of Technology, Pasadena, CA 91125, USA}
\author[0000-0001-5506-9855]{J.~A.~Tomsick}
\affiliation{Space Sciences Laboratory, 7 Gauss Way, University of California, Berkeley, CA 94720-7450, USA}
\author[0000-0003-0870-6465]{B.~M.~Coughenour}
\affiliation{Space Sciences Laboratory, 7 Gauss Way, University of California, Berkeley, CA 94720-7450, USA}
\author[0000-0003-4216-7936]{G~Mastroserio}
\affiliation{Cahill Center for Astronomy and Astrophysics, California Institute of Technology, Pasadena, CA 91125, USA}
\author[0000-0001-9110-2245]{D.~Wik}
\affiliation{Department of Physics and Astronomy, University of Utah, 115 1400 E., Salt Lake City, UT 84112, USA}
\author[0000-0003-2737-5673]{R.~Krivonos}
\affiliation{Space Research Institute of the Russian Academy of Sciences (IKI), Moscow, 117997, Russia}
\author[0000-0002-3850-6651]{A.~D.~Jaodand}
\affiliation{Cahill Center for Astronomy and Astrophysics, California Institute of Technology, Pasadena, CA 91125, USA}
\author[0000-0003-1252-4891]{K.~K.~Madsen}
\affiliation{CRESST and X-ray Astrophysics Laboratory, NASA Goddard Space Flight Center, Greenbelt, MD 20771 USA}
\author{D.~J.~K.~Buisson}

\begin{abstract}


We present an updated catalog of \straycats (a catalog of \nustar stray light observations of X-ray sources) that includes nearly 18 additional months of observations. \straycats\ v2 has an added 53 sequence IDs, 106 rows, and 3 new identified stray light (SL) sources in comparison to the original catalog. The total catalog now has 489 unique sequence IDs, 862 entries, and 83 confirmed \straycats sources. Additionally, we provide new resources for the community to gauge the utility and spectral state of the source in a given observation. We have created long term light curves for each identified SL source using \maxi and \swift/BAT data when available. Further, source extraction regions for 632 identified SL observations were created and are available to the public. In this paper we present an overview of the updated catalog and new resources for each identified \straycats SL source. 

\end{abstract}


\section{Introduction}

The \textit{Nuclear Spectroscopic Telescope Array} (\nustar: \citealt{harrison13}) is the first hard X-ray focusing telescope with a passband from $3-80$ keV. This is achieved through a 10-meter deployed mast that connects the focusing optics to the detectors. 
However, because the mast has an open geometry construct, it is possible for photons to pass through the aperture stop and reach the detectors without interacting with the optics. This is referred to as ``stray light". 
Stray light observations arise for bright X-ray sources 1--4$^{\circ}$ away from the telescope pointing direction \citep{madsen17}, since X-rays  are able pass through the gap between optics and aperture stop and directly illuminate the \nustar focal plane at these off-axis angles.

Recently, the first systematic identification of stray light in \nustar observations was carried out and compiled into a catalog known as \straycats\footnote{https://nustarstraycats.github.io/straycats/} \citep{gref21}. 
The catalog includes serendipitous observations of persistently bright X-ray sources, pulsars, transient neutron star (NS) and black hole (BH) X-ray binaries, as well as intentional stray light observations of very bright targets to reduce telemetry load (e.g., the BH X-ray binary MAXI~J1820+070 during the peak of outburst: \citealt{buisson19}). 
The catalog demonstrated the level of data quality for stray light and data analysis tools that are available via the \nustar community-contributed GitHub page\footnote{https://github.com/NuSTAR/nustar-gen-utils}. 
Additionally, there is an ipython notebook walkthrough on reducing stray light data provided in the repository for new users.

The first science results were recently published in \citet{brumback22} regarding a timing analysis of \nustar \straycats data of the high-mass X-ray binary (HMXB) SMC~X-1. The process of screening stray light data for scientific analysis was detailed therein. 
The source was observed fortuitously at various stages of the binary orbit allowing for a test of the ephemeris and confirming that it remained unchanged. Furthermore, time- and energy-resolved pulse profiles were generated and analyzed, highlighting the variability of pulse profile shape with time in this system. The analysis of the stray light data confirmed and extended the baseline of SMC~X-1's NS spin up rate. This highlights the utility of \straycats for monitoring both short and long term timing behaviors in variable sources.
A forthcoming paper details that stray light for bright hard X-ray sources is viable and well calibrated for spectral analysis above 80~keV since the data are not limited by the reflectivity of the mirrors in the focusing optics (G.~Mastroserio et al.\ 2022, in prep.).

The following sections provide an overview of the work done to update the catalog with nearly 18 additional months of observations and of the supplemental products provided for the public. 

\section{Updated Catalog and Products}
The catalog has been updated to include \nustar observations from the end of the prior release (2020 July 15 to 2022 January 04). \citet{gref21} describes the automated process of identifying potential stray light and the manual process of confirming sources. An additional 1050 observations were examined for stray light utilizing this methodology. Another 53 observations were identified to have stray light illuminating one or both focal plane modules (FPMs). These were added to the existing catalog to produce the updated version presented here. Furthermore, there are 3 new stray light sources that were not in the previous release. These are the NS HMXB 1A 0535+262, NS low-mass X-ray binary (LMXB) 2S 0918-549, and BH LMXB GX 339-4. When \straycats is referenced hereafter it refers to all entries from the prior release and this update.

\begin{figure}[t]
\begin{centering}
\includegraphics[width=0.47\textwidth, trim=20 0 0 0, clip]{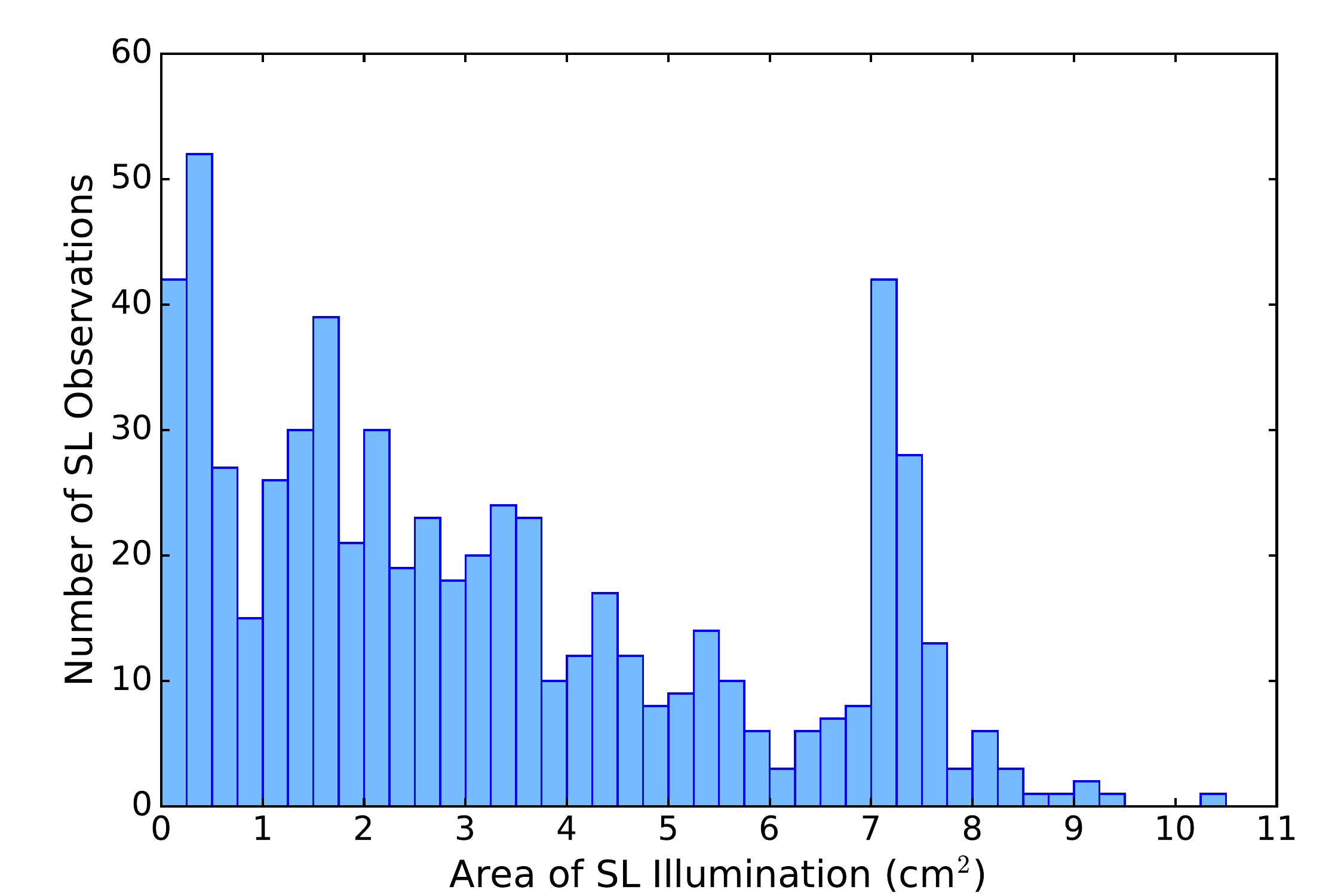}
\caption{Distribution of the size of stray light illumination area of known SL sources within \straycats. Each bin is 0.25 cm$^{2}$. The peak near 7.25 cm$^{2}$ is dominated by the intentional stray light observations of the Crab for calibration purposes and brighter sources to reduce the telescope telemetry load. For reference, the total area per FPM is 16~cm$^{2}$.}
\label{fig:hist}
\end{centering}
\end{figure} 

\subsection{Additional High-level Science Products}
In order to address the needs of the community and streamline the analysis process, we manually generate source extraction regions for all identified stray light sources marked `SL' in \straycats. Images in detector 1 (Det1) coordinates were created from the `make\_det1\_image' wrapper for the identified stray light observations. These were then loaded into DS9 in order to manually create source extraction regions for each entry that are saved in `IMAGE' coordinates. This provides a starting point for the user. These are each labeled according to their given \nustar sequence ID, the focal plane module (FPM) which the pattern is on, and StrayID. For example, the source region for the third entry in the primary version of the catalog is labeled 90201041002A\_StrayCatsI\_3.reg and the hundredth entry in the second version of \straycats is 90701306002A\_StrayCatsII\_100.reg.

These regions were used to extract events for each observation using the `extract\_det1\_events' wrapper in two different energy bands: $3-8$~keV and $8-13$~keV. This upper bound was chosen to ensure that soft X-ray sources are still above the background (see Figure~4 in \citealt{brumback22} for an example of a stray light source spectrum and the associated background component) and sources could be looked at in a uniform manner within the catalog. The user is free to chose different energy bands that suit their science needs.
Additionally, Det1 exposure maps were created via `make\_exposure\_map' wrapper in order to determine the SL illumination area `straylight\_area' tool. Figure \ref{fig:hist} shows a histogram of the size of the stray light illumination area for the known SL sources. The total counts ($\rm C_{tot}$) per energy band and SL area are provided for the user to determine if the observation will be viable for their science needs. Furthermore, the mean count rate per area ($I$) in each band is calculated. Note that these values are not background subtracted.

\begin{figure}[t!]
\begin{centering}
\includegraphics[width=0.48\textwidth, trim=0 0 0 0, clip]{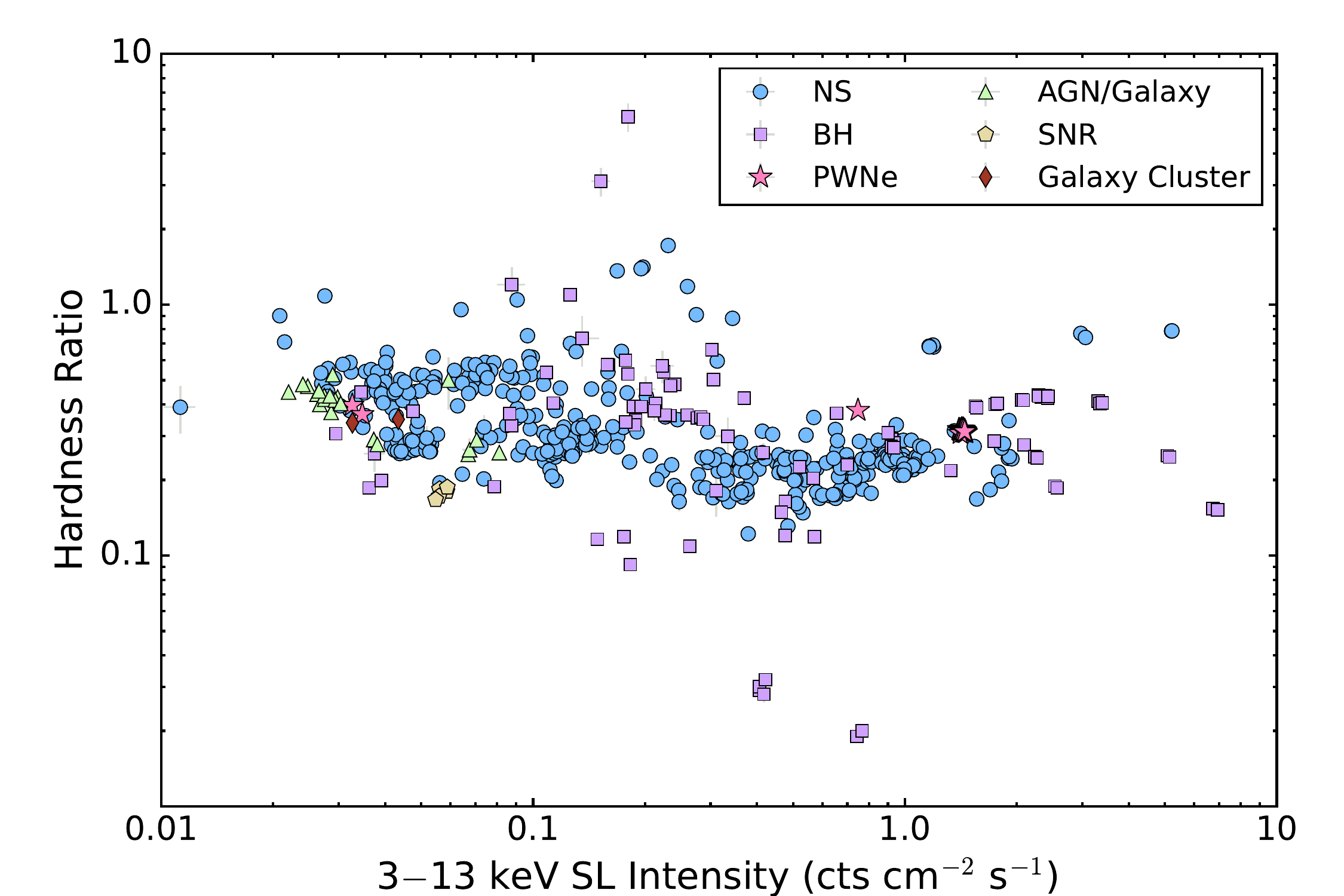}
\caption{Hardness ratio versus $3-13$~keV intensity for all identified SL sources with $1\sigma$ error bars. The markers indicate the classification of the source. The NS and BH sources can be further broken down into HMXBs, LMXBs, and BH candidates (BHC) as denoted in the \straycats FITS file.}
\label{fig:hr}
\end{centering}
\end{figure} 

Many of the X-ray sources in the catalog transition through various spectral states based on whether the spectrum is dominated thermally (spectrally soft) or non-thermally (spectrally hard). 
One straightforward way to gauge the spectral state is to take the ratio of events in different X-ray energy bands to obtain a hardness ratio (HR), although the exact definition of hardness and spectral states differs between BHs (e.g., \citealt{gierlinski06}) and NSs (e.g., \citealt{hasinger89}). We divide the mean count rate per area in the hard X-ray band ($I_{\rm hard}$) by the soft X-ray band ($I_{\rm soft}$) to obtain an approximate HR that can provide a proxy of the spectral state of the source at the time of observation. To demonstrate the general hardness properties of the detected sources, Figure~\ref{fig:hr} shows the HR of the SL sources denoted by classification of the SL source in \straycats versus the $3-13$~keV intensity of the SL source.

\begin{figure}[t]
\begin{centering}
\includegraphics[width=0.48\textwidth, trim=0 0 0 0, clip]{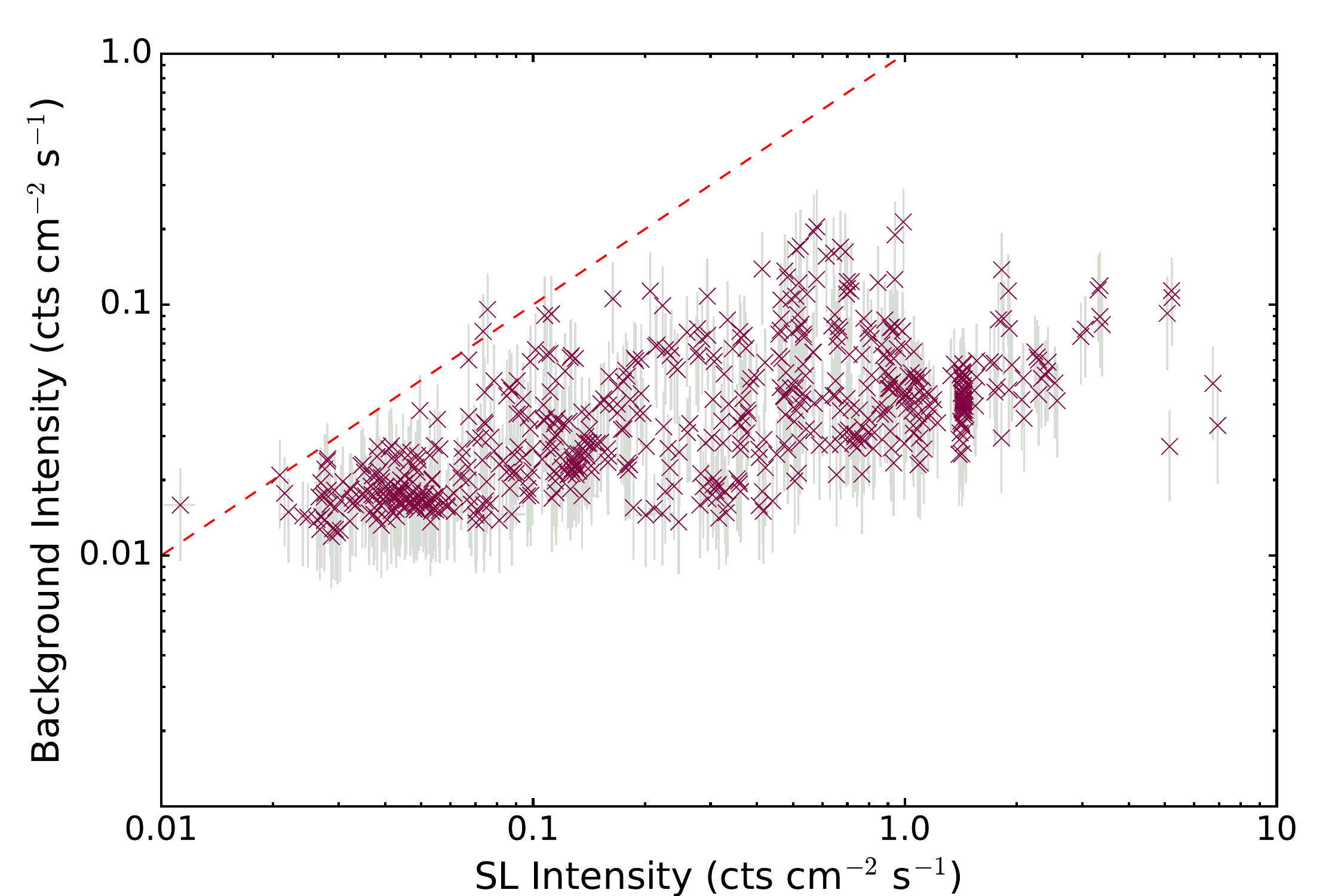}
\caption{The background intensity versus the intensity for identified SL sources in the $3-13$~keV energy band. The error on the background encompasses a 50\% systematic uncertainty while the errors on the SL intensity are at the $1\sigma$ level. The red dashed line indicates where the background is equal to the SL intensity. Many of the sources are detected significantly above the X-ray background estimate with faint SL sources encroaching on the limit of our sensitivity to identify.}
\label{fig:bkg}
\end{centering}
\end{figure} 

\subsection{Background Estimation}
The \nustar background has multiple spectroscopic terms that vary spatially over the focal planes \citep[e.g.,][]{wik14}. Detailed spectroscopic analysis for the \straycats sources will require a complete analysis of the background terms and any transmitted stray light. Here, we estimate the background intensity through a simple masking of the focal plane with the `straylight\_background' tool, which removes the stray light source(s) as well as the flux from the primary target within 3 arcmins of the focused source's location. 
This value was chosen in order to remove any contribution from a focused X-ray bright target, such as a Galactic X-ray binary, but the user is free to choose a different value.
We average over the remaining focal plane area to estimate the average in-band background flux ($I_{\rm bkg}$). 
These values are stored in the updated \straycats FITS table and are provided as notional background values. Given the known variability and intricacies with the \nustar X-ray background, a systematic error at the 50\% level is provided for the $I_{\rm bkg}$ values reported in the \straycats table. We stress that for detailed spectral analysis of stray light data a bespoke model for the background should be used (e.g., \texttt{nuskybgd-py}\footnote{https://github.com/NuSTAR/nuskybgd-py}). 
Users do not need to apply a systematic uncertainty to any science products created through \texttt{nustar-gen-utils}.
An excerpt of the complete table of values is shown in Table \ref{tab:tab1}. A complete fits table is available online. Figure~\ref{fig:bkg} shows the \nustar X-ray background intensity in the $3-13$~keV energy band versus the $3-13$~keV intensity of the SL sources.

\begin{figure}[t!]
\begin{centering}
\includegraphics[width=0.47\textwidth, trim=0 0 0 0, clip]{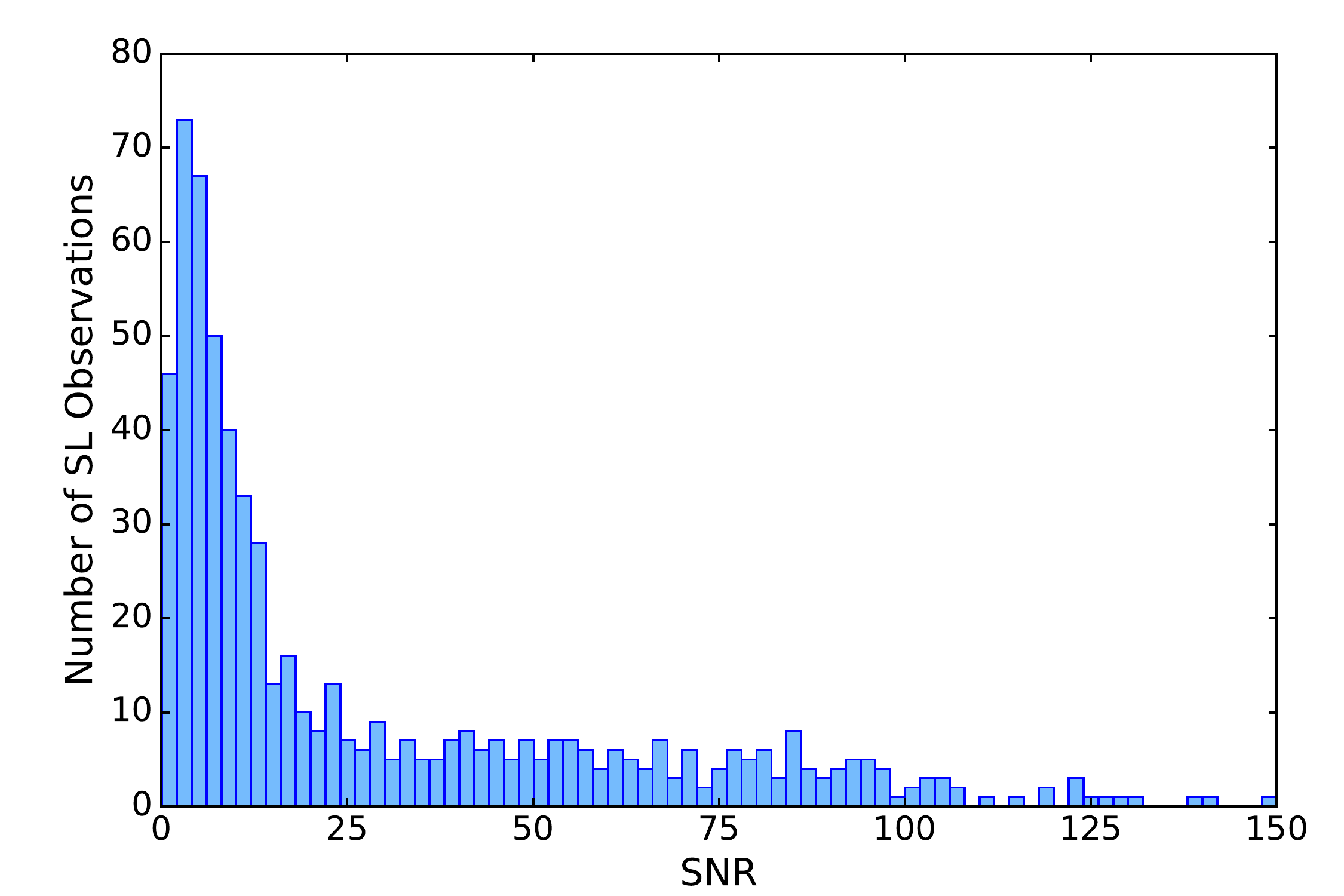}
\caption{Distribution of the SNR for the known SL sources within \straycats. Each bin indicates a step in SNR of 2.}
\label{fig:snr}
\end{centering}
\end{figure}

We use this estimate of the background to compute the effective signal-to-noise ratio (SNR) in the source region (Figure~\ref{fig:snr}). A majority of the high SNR sources are ``intentional" SL observations taken of bright X-ray binaries in outburst (e.g, MAXI J1820+070) or calibration observations of the Crab. The low-SNR tail demonstrates that our methodology can successfully identify SL sources down to a rough approximation of the noise floor. We note that variations in the background and the varying amount of area that a SL source covers on the FoV means that it is not possible to make a strong statement about the completeness of our sample. However, the low-SNR tail demonstrates that we have (likely) identified most \straycats targets where detailed science analyses are possible.

\subsection{Long Term Light Curves}
\begin{figure*}[t]
\begin{centering}
\includegraphics[width=0.99\textwidth, trim=0 0 0 0, clip]{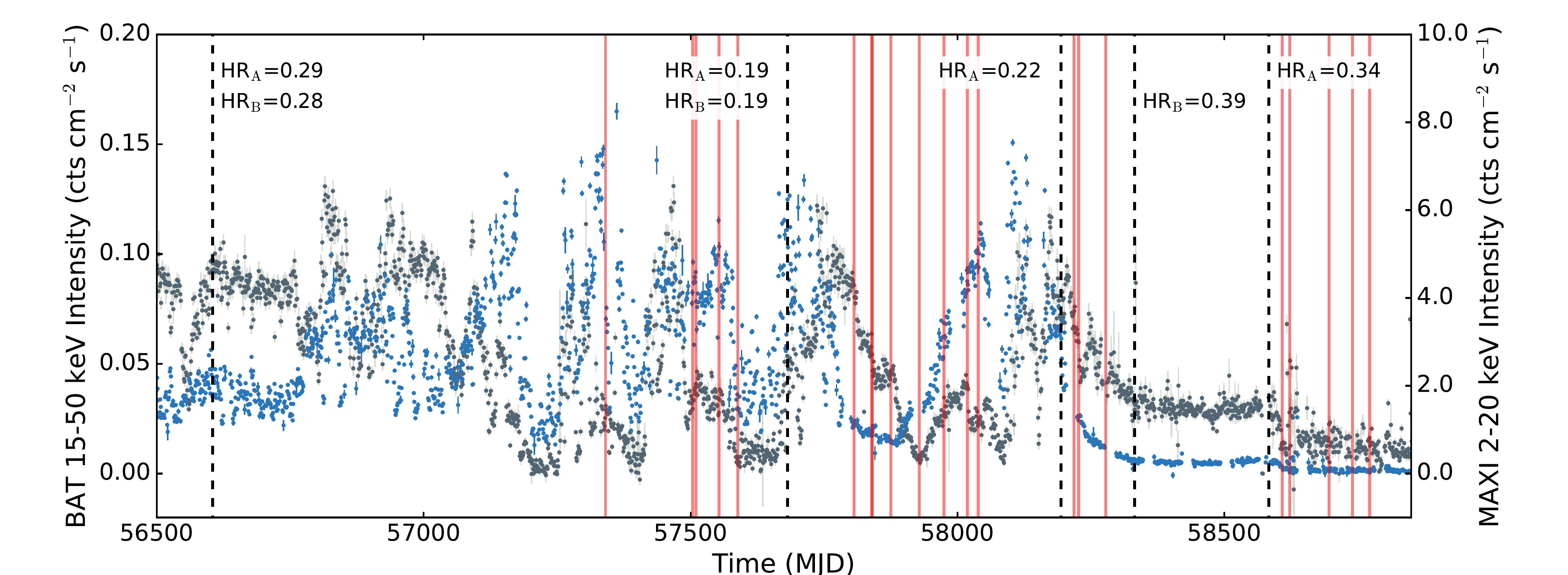}
\includegraphics[width=0.99\textwidth, trim=0 0 0 0, clip]{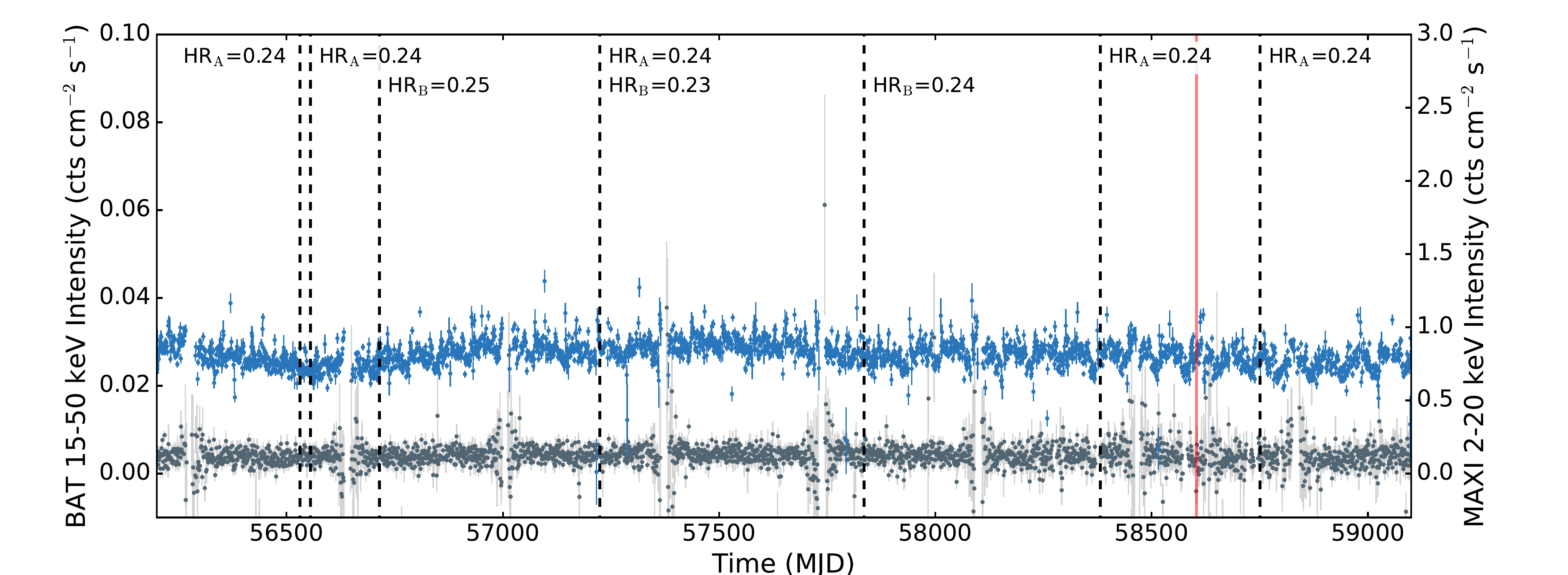}
\caption{The long term MAXI (blue) and Swift/BAT (gray) light curve for the BH LMXB GRS 1915+105 (top) and the NS LMXB GX~9+9 (bottom). The red vertical lines denote focused \nustar observations of the target. The black vertical dashed lines indicate select stray light observations with an illumination pattern larger than 1~cm$^{2}$. The hardness ratio (HR = $I_{\rm hard}$ / $I_{\rm soft}$) for the stray light data is provided for each based on if the stray light pattern occurred on FPMA or FPMB. The error on HR is $\sigma\leq 0.03$ and $\sigma\leq0.02$ for GRS~1915+105 and GX~9+9, respectively.}
\label{fig:lc}
\end{centering}
\end{figure*}

As an additional resource to gauge source state at the time of observing, we use data from the Monitor of All-sky X-ray Image (\maxi: \citealt{maxi}) and \swift/BAT to construct long term light curves for each source when data were available. \nustar focused and stray light observations are marked by a solid red line and dashed black line, respectively.
Figure~\ref{fig:lc} (top) shows an example of a long term light curve for the highly variable BH LMXB GRS 1915+105 similarly to \citet{gref21}.
However, we also show the HR value for each stray light observation with an illumination pattern $\geq1~{\rm cm}^{2}$ per focal plane module (FPM). 
Examples of the \nustar light curves of GRS 1915+105 during the stray light observations are provided in \citet{gref21}.

As an additional example, we show the long term light curve of the persistently accreting NS LMXB GX~9+9 in Figure~\ref{fig:lc} (bottom). The source is less variable than the BH LMXB GRS~1915+105 and classified as an `atoll' source based upon characteristic island like shapes that are traced out on X-ray color-color and hardness-intensity diagrams \citep{hasinger89}. 
The source is part of a subclass of atolls that have only been observed to trace out the banana branch. Even though the source intensity varies from observation to observation, the HR remains roughly constant, which is inline with the overall spectral state remaining in the banana branch. 

\section{Conclusion}
We have updated \straycats to include nearly 18 additional months of observations and provide  resources to the community to streamline the process of analyzing \nustar stray light data. Source extraction regions were created for 632 identified stray light source observations. We provide information on the total counts and intensity in two energy bands ($3-8$~keV and $8-13$~keV) for each of the identified stray light observations, as well as the HR and proxy for the background intensity. Furthermore, for each SL source, we have created long-term \maxi and \swift/BAT light curves when data were available. We denote times when \nustar focused and stray light observations occurred. These products and updated catalog are available on the \straycats GitHub page\footnote{https://nustarstraycats.github.io/straycats/} and are intended to help the community in assessing if a given stray light observation will prove fruitful for their science objectives. \\

{\it Acknowledgements:} Support for this work was provided by NASA through the NASA Hubble Fellowship grant \#HST-HF2-51440.001 awarded by the Space Telescope Science Institute, which is operated by the Association of Universities for Research in Astronomy, Incorporated, under NASA contract NAS5-26555. This work was additionally supported by the National Aeronautics and Space Administration (NASA) under grant number 80NSSC19K1023 issued through the NNH18ZDA001N Astrophysics Data Analysis Program (ADAP). This research has made use of the \nustar Data Analysis Software (NuSTARDAS) jointly developed by the ASI Science Data Center (ASDC, Italy) and the California Institute of Technology (USA).

\begin{table*}[]
\caption{Excerpt from the updated \straycats table that contains additional information on the identified stray light sources}
\label{tab:tab1}

\begin{center}
\begin{tabular}{lcccccccc}
\hline

StrayID & Area & $\rm C_{ tot,\ soft}$ & $\rm C_{tot,\ hard}$ & HR & $I_{\rm soft}$ & $I_{\rm hard}$ & $I_{\rm bkg,\ soft}$ & $I_{\rm bkg,\ hard}$\\
\hline

StrayCatsI\_2 & 0.69 & $ 996 \pm 32 $  & $509 \pm 23$ & $0.51\pm0.03$ & $2.64\pm0.08$ & $ 1.35 \pm 0.06$ & $ 1.1 \pm 0.6$ & $ 0.5\pm 0.3$ \\ 
StrayCatsI\_3 & 4.31 & $ 4121 \pm 64 $  & $2024 \pm 45 $ & $0.49 \pm 0.01$ & $2.48\pm0.04$ & $ 1.22 \pm 0.03$ & $ 1.4 \pm 0.7$ & $ 0.7\pm 0.4$ \\ 
StrayCatsI\_4 & 7.08 & $ 29469 \pm 172 $  & $15015 \pm 123 $ & $0.51 \pm 0.01$ & $5.89\pm 0.03$ & $ 3.0 \pm 0.02$ & $ 1.6 \pm 0.8$ & $ 0.9\pm 0.4$ \\ 
StrayCatsI\_5 & 6.97 & $ 13003 \pm 114 $  & $6851 \pm 83$ & $0.53 \pm 0.01$ & $ 3.2\pm 0.03$ & $ 1.68 \pm 0.02$ & $ 1.0 \pm 0.5$ & $ 0.5\pm 0.3$ \\ 
StrayCatsI\_8 & 0.66 & $ 3159 \pm 56 $  & $1666 \pm 41$ & $0.53 \pm 0.02$ & $ 6.5 \pm 0.1$ & $ 3.4\pm 0.08$ & $ 1.2\pm 0.6$ & $0.6\pm 0.3$ \\
\multicolumn{9}{c}{...}\\
StrayCatsII\_99 & 1.30 & $ 6228 \pm 79 $  & $1099 \pm 33$ & $0.18 \pm 0.01$ & $ 27.5 \pm 0.4$ & $ 4.9 \pm 0.2$ & $ 2.2 \pm 1.1$ & $ 0.6 \pm 0.3$ \\ 
StrayCatsII\_100 & 7.86 & $ 283917 \pm 533$  & $59356 \pm 244 $ & $0.209 \pm 0.001$ & $ 80.1 \pm 0.2$ & $ 16.74 \pm 0.07$ & $ 3.6 \pm 1.8$ & $ 1.1 \pm 0.6$ \\ 
StrayCatsII\_101 & 8.93 & $ 326386 \pm 571 $  & $68319 \pm 261 $ & $0.209 \pm 0.001$ & $ 81.9 \pm 0.1$ & $ 17.15 \pm 0.07$ & $ 4.0 \pm 2.0$ & $ 1.0 \pm 0.5$ \\ 
StrayCatsII\_104 & 0.44 & $ 5948 \pm 77 $  & $1367 \pm 37 $ & $0.23 \pm 0.01$ & $ 56.9 \pm 0.7$ & $ 13.1 \pm 0.4$ & $ 8.6 \pm 4.3$ & $ 3.0 \pm 1.5$ \\ 
StrayCatsII\_105 & 3.49 & $ 6682 \pm 82 $  & $1113 \pm 33 $ & $0.17 \pm 0.01$ & $ 4.69 \pm 0.06$ & $ 0.78 \pm 0.02$ & $ 1.1 \pm 0.5$ & $ 0.5 \pm 0.2$ \\ 
\hline 

\end{tabular}
\end{center}

\medskip
Note.---  The StrayID refers to the row number in \straycats. The illuminating area in units of cm$^{2}$ are given for each observation. The total events ($\rm C_{tot}$) and count rate per unit area ($I$ [10$^{-2}$~counts~cm$^{-2}$~s$^{-1}$]) are provided for two energy bands: $3-8$~keV (soft) and $8-13$~keV (hard). The hardness ratio (HR) is the intensity in the hard energy band divided by the intensity in the soft energy band. We also provide the intensity of the background in the units of 10$^{-2}$~counts~cm$^{-2}$~s$^{-1}$ in the two energy bands for reference. All errors are given at the $1\sigma$ uncertainty level except for $I_{\rm bkg}$, which has a 50\% systematic uncertainty given the known variability and intricacies in the \nustar X-ray background.
\end{table*} 


\begin{thebibliography}{}
\bibitem[Brumback et. al.(2022)]{brumback22} Brumback, M. C., Grefenstette, B. G., Buisson, D. J. K., et al.\ 2022, ApJ, 926, 187
\bibitem[Buisson et al.(2019)]{buisson19} Buisson, D. J. K., Fabian, A. C., Barret, D., et al.\ 2019, MNRAS, 490, 1315
\bibitem[Gierli\'nksi \& Newton(2006)]{gierlinski06} Gierli\'nksi, M., \& Newton, J.\ 2006, MNRAS, 370, 837
\bibitem[Grefenstette et. al.(2021)]{gref21} Grefenstette, B. G., Ludlam, R. M., Thompson, E. T., et al.\ 2021, ApJ, 909, 30
\bibitem[Harrison et al.(2013)]{harrison13}Harrison, F. A., Craig, W. W., Christensen, F. E., et al.\ 2013, \apj, 770, 103
\bibitem[Hasinger \& van der Klis(1989)]{hasinger89}Hasinger, G., \& van der Klis, M.\ 1989, A\&A, 225, 79
\bibitem[Madsen et al.(2017)]{madsen17}Madsen, K. K., Christensen, F. E., Craig, W. W., et al.\ 2017a, Journal of Astronomical Telescopes, Instruments, and Systems, 3, 044003
\bibitem[Matsuoka et al.(2009)]{maxi}Matsuoka, M., Kawasaki, K., Ueno, S., et al.\ 2009, PASJ, 61, 999
\bibitem[Wik et al.(2014)]{wik14} Wik, D. R., Hornstrup, A., Molendi, S., et al.\ 2014, ApJ, 792, 48
\end{thebibliography}
\end{document}